\documentclass[11pt,twoside]{article}
\usepackage{asp2014}

\aspSuppressVolSlug
\resetcounters

\begin{document}

\title{Space-Time Coverage in the VO Registry}

\author{Markus~Demleitner$^1$}
\affil{$^1$Universit{\"a}t Heidelberg, Zentrum f{\"u}r Astronomie,
Heidelberg, Germany \email{msdemlei@ari.uni-heidelberg.de}}

\paperauthor{Markus~Demleitner}{}{}{Universit{\"a}t Heidelberg}
{Zentrum f\"ur Astronomie}{Heidelberg}{}{69221}{Germany}



  
\begin{abstract}
With VODataService 1.2, service providers in the Virtual Observatory
(VO) have a reasonably straightforward way to declare where in space,
time, and spectrum the data within a resource (i.e., service or data
collection) lie.  Here, we discuss the the mechanism and design
choices, current limitations (e.g., regarding non-electromagnetic or
solar system resources) as well as ways to overcome them.  We also show
how users and clients can already run queries against resoure coverage
using a scheme that is exprected to become part of RegTAP 1.2 (or a
separate standard).  We conclude with an ardent plea to all resource
creators to provide STC metadata {\textendash} only wide adoption will
make this facility useful.

\end{abstract}

\section{Introduction}

Last year, \citet{std:stcReg} proposed a roadmap for how to finally
provide means for resource discovery using space, time, and spectral
constraints in the Virtual Observatory.  This is not only useful in
itself, it is also a precondition for efficient blind discovery, i.e.,
finding research artefacts not by previous knowlegde of particular data
collections (``2MASS magnitudes'') but by physical properties (``fluxes
around $2\,\mu \rm m$ for red giants in the LMC'').

The roadmap proposed five concrete steps to be taken, of which three are
rather technical in nature.  The remaining two were an update to the
VODataService standard that enables data providers to declare the
coverage of their resources, and an update to the resource discovery
protocol RegTAP that lets clients use these coverages in their
queries.

In late 2019, a Proposed Recommendation for VODataService 1.2 is
available \citep{pr:VODS12}, and the ideas for RegTAP evolution laid out
in the roadmap have a prototype implementation.  This contribution reports
on both.

\section{Declaring Space-Time Coverage}

To declare where data within a resource is located in the VO phase
space, VODataService 1.2 defines three new child elements of the
pre-existing \texttt{coverage} element:

\begin{itemize}
\item \texttt{spatial} -- zero or one allowed.  Contains an ASCII MOC
\citep{2014ivoa.spec.0602F}, written in the ICRS.
\item \texttt{temporal} -- zero or more allowed.  Contains a
space-separated pair of floats giving an MJD interval.
\item \texttt{spectral} -- zero or more allowed.  Contains a
space-separated pair of floats giving an energy interval observed.
Energies are given in Joules. 
\end{itemize}

Diverging from the original roadmap, spectral coverage is now represented
with energy, which which not only is medium-independent (the standard
choice in the VO, wavelength, is not) but also retains its meaning
across different messengers (e.g., neutrinos or charged particles in the
solar system).  The choice of Joule as the unit is somewhat arbitrary
but at least does not favour astronomers working in any particular part
of the spectrum, and it emphasises that user interfaces have to provide
elements adapting to their users' preferred units.

While we do not expect the resulting differences to matter much in
discovery of extrasolar data, all coordinates should be given as
observed from the solar system barycenter, and times should use TDB.

With this, the coverage for a resource of a longer observation campaign
(here, the Palomar-Leyden Trojan surveys) could look like this:

{\small
\begin{verbatim}
<coverage>
  <spatial>3/282,410
    4/40,323,326,329,332,387,390,396,648-650,1083,1085,1087,
    1101-1103,1123,1125,1132-1134,1136,1138-1139,1144,1146-1147,
    ...</spatial>
  <temporal>37190 37250</temporal>
  <temporal>38776 38802</temporal>
  <temporal>41022 41107</temporal>
  <temporal>41387 41409</temporal>
  <temporal>41936 41979</temporal>
  <temporal>43416 43454</temporal>
  <spectral>3.01e-19 6.02e-19</spectral>
  <waveband>Optical</waveband>
</coverage>
\end{verbatim}}

While the standard is agnostic on the resolutions on all axes, as a
good compromise between the desires for compact registry records and
specificity in discovery we currently recommend a maximal level of 6 in
the spatial coverage.

\section{Discovery Using Space-Time Metadata}

To enable discovery on these coverages, the roadmap suggested an
extension to RegTAP that, adapted for the change of spectral modelling,
is already implemented on the RegTAP service at
\url{http://dc.g-vo.org/tap}.  Three tables are added:

\begin{itemize}
\item \texttt{rr.stc\_spatial} with columns \texttt{ivoid},
\texttt{coverage}, \texttt{ref\_system\_name}, where
\texttt{coverage} contains a
MOC-typed geometry which so far only supports the CONTAINS function with
POINTs.
\texttt{ref\_system\_name} is NULL for ICRS.
Non-celestial data will have non-empty values in that column.
\item \texttt{rr.stc\_temporal} with columns \texttt{ivoid},
\texttt{time\_start}, \texttt{time\_end} (with MJD values).
\item \texttt{rr.stc\_spectral} with columns \texttt{ivoid},
\texttt{spectral\_start}, \texttt{spectral\_end} (in Joules of
energy).
\end{itemize}

\articlefigure[height=5cm]{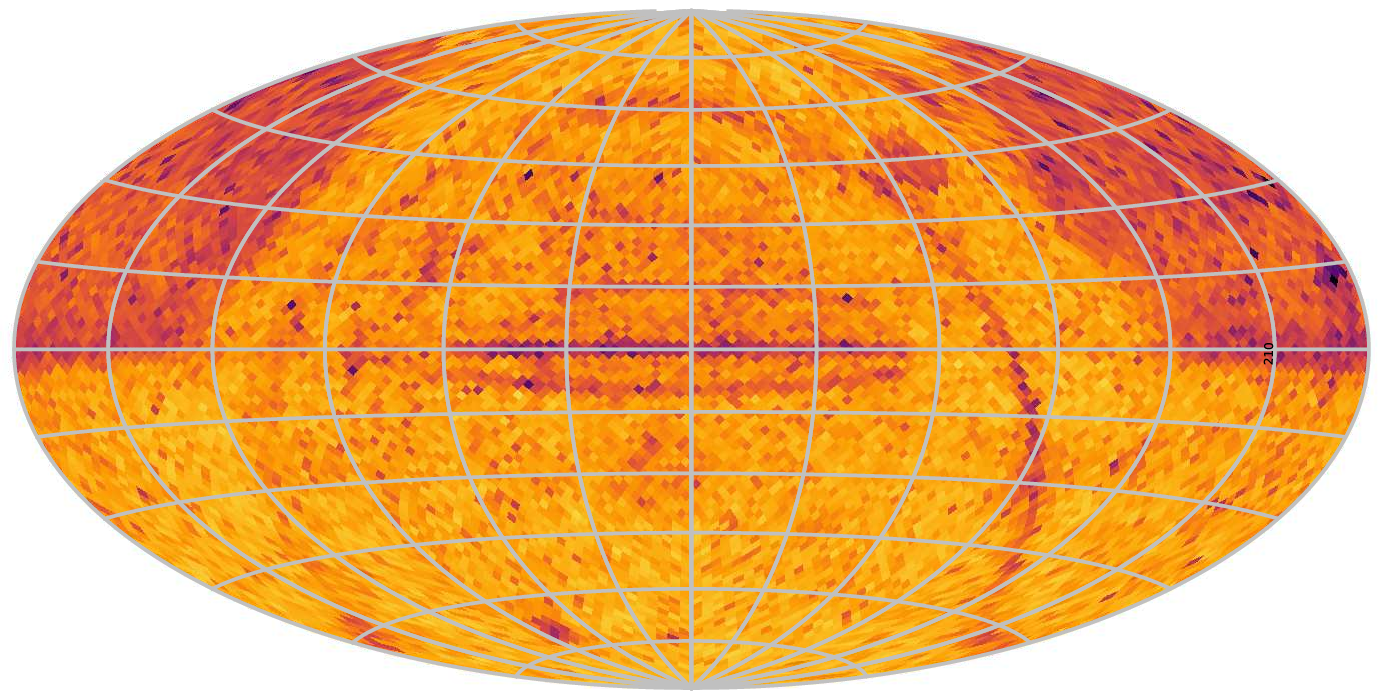}{fig:p22-cov}{A density plot of
resources in the VO Registry with spatial footprints.  Equatorial
system, Aitoff projection.}

A straightforward case to illustrate the use of these tables would be
the discovery of data for messengers around the rest mass of electrons for
the center of the LMC.  In ADQL, this could be written as:

{\small
\begin{verbatim}
SELECT ivoid
FROM rr.stc_spatial
NATURAL JOIN rr.stc_temporal
WHERE
  1=CONTAINS(POINT(80.9, -69.8), coverage)
  AND ref_system_name IS NULL
  AND ivo_interval_overlaps(500, 550,
    in_unit(spectral_start, 'keV'), 
    in_unit(spectral_end, 'keV'))
\end{verbatim}}

The new syntax for geometry construction and the \verb|in_unit| extension
are from ADQL 2.1.  The \verb|ivo_interval_overlaps| is a simple user
defined function also proposed in the Roadmap.

Another, perhaps only bibliometrically useful, example is the creation
of a density map of VO resources; with an ADQL extension proprietary to
DaCHS \citep{2014A+C.....7...27D} that generates integer series, here is
a query to produce a HEALPix map (12287 is the largest HEALPix index on
level 5):

{\small
\begin{verbatim}
SELECT hpx, COUNT(*)
FROM generate_series(0, 12287) AS hpx
JOIN rr.stc_spatial
ON (1=CONTAINS(ivo_healpix_center(5, hpx), coverage))
\end{verbatim}}

Fig.~\ref{fig:p22-cov} shows a visualisation of this data prepared using
TOPCAT \citep{2005ASPC..347...29T}.  We note that the high impact
projects like SDSS and Kepler had on astronomy shows in this plot.

\section{Further Directions}

While the core functionality envisioned by the roadmap is now available,
the scheme can and will be evolved.  Issues we would like to see
addressed include:

\textbf{Solar system data:} Right now, the spatial coverage is always given
in the ICRS.  We believe this is adequate for celestial (in the sense
of: outside of the solar system) data.  It is clearly rather useless for
resources covering solar system objects and phenomena.  At this point we
hope to cover these wider use cases with properly defined terms in
\texttt{ref\_system\_name}.

\textbf{Redshift/Distance:} The first version of the data model for
space-time coordinate offered redshift as
another axis, and one could make a point that distance is part of the
spatial location.  It is, however, hard to define a distance measure that works
naturally from solar system science to cosmology.  Since the number of
resources that can even sensibly give distances appears to be low, we
have postponed this question for now.

\textbf{Non-EM messengers:} There is no way to look for ``Neutrino'' or
``Gravitational Wave'' by this scheme yet (but GeV-neutrinos can be told
from from eV-neutrinos).  It seems at this point that no new VOResource
feature is strictly required to solve this.  The most straightforward
solution would be to expand the waveband vocabulary to new messengers,
even though the element name ``waveband'' might then appear a slight
misnomer.  An alternative avoiding this might be to reserve terms given
in subject for this purpose.

\section{Takeup}

Unfortunately, not many registry resources declare their STC coverage
yet.  As of 2019-09-23, the VO registry contained 15131 resources with
\textit{spatial} coverage (but over 90\% of these are actually havested
from their footprint URLs and thus do not give inline coverages yet), 80
resources with \textit{temporal} coverage (originating from 5
authorities), and 75 resources with \textit{spectral} coverage
(originating from 4 authorities)

At least on temporal and spectral, this is clearly not enough to even
start basing user interfaces on it.  

We therefore conclude with an ardent plea to publishers of VO services
to add coverage information to their registry records.

\bibliography{P2.2}


\end{document}